# Electron mobility of SnO$_2$ from first principles

Amanda Wang[1,a], Kyle Bushick[1], Nick Pant[1,2], Woncheol Lee[3], Xiao Zhang[1], Joshua Leveillee[4,5], Feliciano Giustino[4,5], Samuel Poncé[6,7], Emmanouil Kioupakis[1]

[1] Department of Materials Science and Engineering, University of Michigan, Ann Arbor, Michigan 48109, USA

[2] Applied Physics Program, University of Michigan, Ann Arbor, Michigan 48109, USA

[3] Department of Electrical and Computer Engineering, University of Michigan, Ann Arbor, Michigan 48109, USA

[4] Oden Institute for Computational Engineering and Sciences, The University of Texas at Austin, Austin, Texas 78712, USA

[5] Department of Physics, The University of Texas at Austin, Austin, Texas 78712, USA

[6] European Theoretical Spectroscopy Facility, Institute of Condensed Matter and Nanosciences, Université catholique de Louvain, Chemin des Étoiles 8, B-1348 Louvain-la-Neuve, Belgium

[7] WEL Research Institute, Avenue Pasteur, 6, 1300 Wavre, Belgium

[a] Corresponding author: amandw@umich.edu

The transparent conducting oxide SnO$_2$ is a wide bandgap semiconductor that is easily n-type doped and widely used in various electronic and optoelectronic applications. Experimental reports of the electron mobility of this material vary widely depending on the growth conditions and doping concentrations. In this work, we calculate the electron mobility of SnO$_2$ from first principles to examine the temperature- and doping-concentration dependence, and to elucidate the scattering mechanisms that limit transport. We include both electron-phonon scattering and electron-ionized impurity scattering to accurately model scattering in a doped semiconductor. We find a strongly anisotropic mobility that favors transport in the direction parallel to the *c*-axis. At room temperature and intrinsic carrier concentrations, the low-energy polar-optical phonon modes dominate scattering, while ionized-impurity scattering dominates above $10^{18}$ cm$^{-3}$.

Rutile SnO$_2$ is a transparent conductor that finds applications in devices such as photovoltaics, sensors, and transistors.[1–3] The ease of n-type doping by unintentionally incorporated hydrogen or through impurities (F, Sb) enables free-electron concentrations in the $10^{16}$-$10^{20}$ cm$^{-3}$ range.[4] Since the first synthesis of SnO$_2$ single crystals,[5] numerous experimental studies have reported a wide range of electron mobility values.[6–12] In contrast, first-principles computational studies of electron transport in SnO$_2$, which enable the determination of the theoretical upper mobility limit and the differentiation of the various electron-scattering mechanisms, have emerged only recently. These existing studies use models and approximations, such as the Fröhlich and acoustic deformation potential (ADP) scattering models and the relaxation time approximation (RTA), to calculate electron mobility.[13,14] While they reduce the computational cost needed to quantify electron scattering and its effect on transport, they have limited accuracy.[15,16]

In contrast to models, maximally localized Wannier functions (MLWFs) provide an efficient and predictive computational framework to interpolate the key quantities of carrier transport (electron energies, phonon frequencies, scattering matrix elements) determined from first principles to fine Brillouin-zone (BZ) sampling grids.[17–19] MLWFs exploit the spatial localization of electron and phonon wave-functions in real space to interpolate these quantities to arbitrary wave-vectors. The MLWFs are constructed using first-principles calculation on a coarse BZ grid and retain first-principles accuracy when interpolating to the fine grids needed to capture electron-scattering processes. The interpolated quantities are then used to iteratively solve the Boltzmann transport equation (BTE) and calculate mobility values.[20] To make a meaningful comparison to experiment, mobility calculations need to include the two dominant forms of scattering in heavily-doped semiconductors: electron-phonon scattering and electron-ionized impurity scattering. The phonon-limited mobility (intrinsic mobility) represents the upper limit in defect-free semiconductors. However, because $SnO_2$ is often doped to high carrier concentrations and has shallow donor energies,[4] the resulting ionized dopants also constitute a significant source of scattering. While doping increases the carrier concentration and hence the conductivity, it is important to also consider how mobility is adversely affected by the resulting increased ionized-impurity concentration. By including the scattering rates from both electron-phonon and electron-ionized impurity interactions, the combined effect of the two scattering mechanisms can be captured in the solution of the BTE.[16]

In this work, we calculate the electron mobility in rutile $SnO_2$ as a function of temperature and doping concentration from first principles, accounting for scattering by both phonons and ionized impurities. We study the temperature dependence of the phonon-limited mobility and identify the polar-optical phonons as the dominant scattering modes, finding that at room temperature, the low-energy polar-optical phonons around 30 meV contribute to 82% of the scattering. The total mobility, limited by both phonons and ionized impurities, decreases with increasing carrier concentration due to increasing scattering from ionized dopants. The mobility is anisotropic and highest along the $c$-axis, due to lower electron effective mass and reduced scattering along this direction. We also calculate the Hall mobility and find good agreement with the highest experimentally measured room-temperature mobility values of 260 cm$^2$/V·s.[9]

To calculate the electron mobility of $SnO_2$, we iteratively solve the Boltzmann transport equation using material parameters obtained from density functional theory (DFT), density functional perturbation theory (DFPT), and many-body perturbation theory ($G_0W_0$ method) as summarized below. Specifics of calculation parameters are provided in the supplementary material.

We used Quantum ESPRESSO[21,22] to carry out DFT structural-relaxation calculations within the local density approximation (LDA). The calculated lattice parameters and other material properties of $SnO_2$ are in good agreement with previous experimental and theoretical studies (Table I). The phonons and electron-phonon matrix elements were calculated using DFPT.[23] The phonon dispersion and frequencies at the Γ point are shown in Figure S2 and Table SI.[24,25]

To improve the description of the band structure, we used the $G_0W_0$ method within BerkeleyGW[26,27] to calculate quasiparticle corrections to energies of the electronic states.[28–30] The quasiparticle corrections improve the severely underestimated LDA bandgap (0.749 eV) to a value of 3.182 eV that is in better agreement with experiment (Table I). The $G_0W_0$ bandgap is still lower than the experimental value of 3.59 eV,[31] but it is the curvature of the bands that affects the calculation of transport properties. The electron effective masses evaluated from the $G_0W_0$ quasiparticle band structure, depicted in Figure 1, are in good agreement with both experiment and previous theoretical studies (Table I).[31–36]

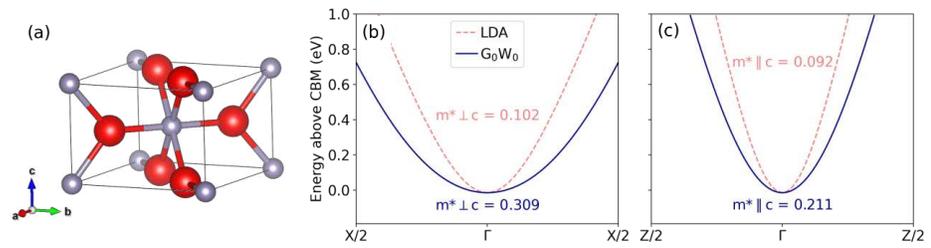

Figure 1 (a) Unit cell of rutile $SnO_2$. The tin atoms are shown in gray and the oxygen atoms in red. (b, c) Conduction band energies referenced to the conduction band minimum (CBM) along high-symmetry paths X-Γ-X and Z-Γ-Z, with the effective masses along each direction labeled.

Table I Structural, electronic, and optical properties of rutile $SnO_2$, as calculated in this work and compared to previous theoretical and experimental studies. The structural parameters were obtained through a relaxation with the LDA functional and valence pseudopotentials. Semicore pseudopotentials and experimental lattice parameters were used for $G_0W_0$ quasiparticle corrections to the band structure. * denotes values from quasiparticle $G_0W_0$ calculations.

|  | a (Å) | c (Å) | u | $E_g$ (eV) | m*⊥ c | m* ∥ c | $\varepsilon_0^\perp$ | $\varepsilon_0^\parallel$ | $\varepsilon_\infty^\perp$ | $\varepsilon_\infty^\parallel$ |
|---|---|---|---|---|---|---|---|---|---|---|
| This work | 4.788 | 3.248 | 0.307 | 3.182* | 0.309* | 0.211* | 14.73 | 10.65 | 5.05 | 5.31 |
| Previous theory | 4.727[32] | 3.200[32] | 0.306[32] | 3.65*[32] | 0.26*[32] | 0.21*[32] | 12[34] | 7.0[34] | 3.7[34] | 3.9[34] |
| Experiment | 4.740[33] | 3.190[33] | 0.306[33] | 3.59[31] | 0.299[35] | 0.234[35] | 13.5[36] | 9.58[36] | 3.785[36] | 4.175[36] |

We used the EPW code[19] to calculate the electron mobility. The quasiparticle energies, phonon frequencies, and electron-phonon coupling matrix elements were calculated on an 8×8×12 Brillouin zone sampling grid and interpolated to finer grids using the maximally localized Wannier function method as implemented in Wannier90.[17,37] We include both dipole and quadrupole corrections to the interpolation of the long-range components of the electron-phonon matrix elements.[38–40] The quadrupole tensor (Table SII) was calculated using ABINIT[41–43] with LDA pseudopotentials without nonlinear core corrections,[44] while the dipole corrections are handled by EPW.[19] The interpolated quantities were used to solve the linearized BTE to calculate the phonon-limited and ionized-impurity-limited[16] electron drift and Hall[45] mobilities as a function of temperature and carrier concentration. The equations used for the mobility calculations are reported in section S2. We assume complete donor ionization and set the carrier concentration equal to the ionized-impurity concentration. To examine the temperature dependence, we set the carrier concentration at $10^{17}$ cm$^{-3}$ and calculated the mobilities from 100 – 600 K. We also investigated the carrier-concentration dependence over the $10^{16} - 10^{20}$ cm$^{-3}$ range at 300 K.

We find that the room-temperature electron drift mobility of SnO$_2$, considering the combined effects of both phonon and ionized-impurity scattering, for a carrier concentration of $10^{17}$ cm$^{-3}$ is 373 cm$^2$/V·s in the direction parallel to the *c*-axis and 228 cm$^2$/V·s perpendicular to the *c*-axis. This anisotropy reflects the difference in effective masses, where m$^*$ is heavier in the perpendicular direction. The anisotropy in the mobilities is stronger than in the effective masses (1.64 compared to 1.46), indicating that, in addition to a directional dependence in the band curvature, there is also anisotropy in the scattering rates.

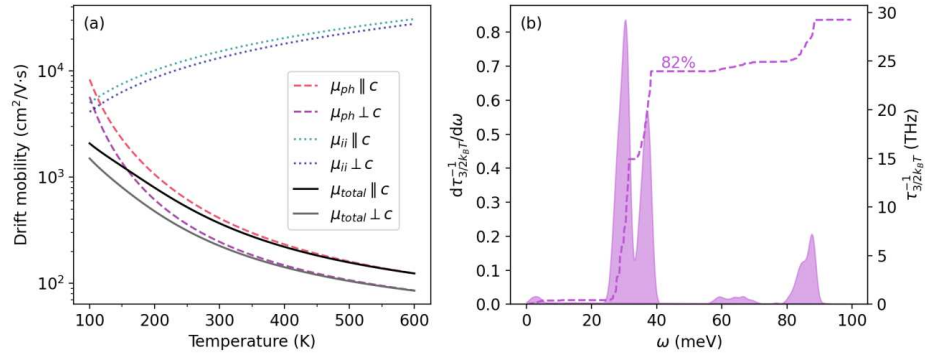

*Figure 2 (a) Electron drift mobility of rutile SnO$_2$ along ∥ c and ⊥ c as a function of temperature at a carrier concentration of $10^{17}$ cm$^{-3}$. (b) Spectral decomposition of the angularly averaged electron-phonon scattering rates by phonon energy at 300 K, for carriers $3k_bT/2$ from the band edge. The dashed line is the cumulative integral of the scattering rates and 82% represents the contribution from the lower-energy polar-optical phonon modes.*

We next examine the temperature dependence of the drift mobility and its decomposition in terms of scattering by phonons and ionized impurities (Figure 2). The phonon-limited mobility decreases with increasing temperature, as expected due to higher phonon occupations at elevated temperatures. We fit the temperature dependence of the phonon-limited mobility according to[46]:

$$\mu_{\text{ph}}(T) = \left(\frac{1}{\mu_{\text{low}}} e^{-T_{\text{low}}/T} + \frac{1}{\mu_{\text{high}}} e^{-T_{\text{high}}/T}\right)^{-1}. \quad (1)$$

The first term on the right-hand side represents lower-energy phonon modes that dominate scattering at lower temperatures, while the second term represents higher-energy phonon modes that dominate at higher temperatures. In polar semiconductors where macroscopic electric fields created by longitudinal-optical phonons can couple to electrons, polar-optical phonons are often the dominant source of phonon scattering.[15] Our calculations confirm that the polar-optical modes contribute the most to limiting the mobility. The $T_{\text{low}}$ and $T_{\text{high}}$ that we fit (Table SIII) correspond to the energies of polar-optical E$_u$ modes with approximate frequencies of 30-40 meV and 80 meV, respectively. These same polar-optical modes also have the largest mode-resolved electron self-energies (Figure S3). The spectral decomposition of scattering rates in Figure 2b show that at room temperature, the lower-energy $E_u$ optical phonon modes with energies of 30-40 meV dominate scattering, contributing to 82% of the scattering rate, as the occupations of the higher-energy $E_u$ mode is low (4-6%). Because the dominant longitudinal-optical modes are well-described by dipole interactions, the inclusion of quadrupole interactions changes the mobilities by less than 1% (Section S5).

We also found that the ionized-impurity-limited mobility increases with increasing temperature. This is because carriers with higher kinetic energy are less affected by impurities at higher temperatures. The temperature dependence of the ionized-impurity-limited mobility is characterized by a power-law model:

$$\mu_{\text{ii}}(T) = \mu_0 \left(\frac{T}{300}\right)^\alpha \quad (2)$$

where we normalize temperature and set $\mu_0$ to the ionized-impurity-limited mobility at 300 K. For SnO$_2$, we find that α is close to 1 for both directions (Table SIII), indicating that there is a nearly linear dependence of the ionized-impurity-limited mobility on temperature.

The total drift mobility, which includes the effects of both phonon and ionized-impurity scattering, is dominated by phonon scattering, as seen by its decreasing value with increasing temperature. We characterize the temperature dependence of the total mobility by combining the previous two equations according to a Matthiessen's rule-like expression[47]:

$$\mu_{\text{total}}(T) = \left(\frac{1}{\mu_{\text{low}}} e^{-T_{\text{low}}/T} + \frac{1}{\mu_{\text{high}}} e^{-T_{\text{high}}/T} + \frac{1}{\mu_0}\left(\frac{T}{300}\right)^{-\alpha}\right)^{-1}, \quad (3)$$

where the fitted parameters are listed in Table SIII. We find that for a moderate dopant concentration of $10^{17}$ cm$^{-3}$, ionized-impurity scattering is weak and electron-phonon scattering dominates the mobility. Furthermore, Matthiessen's rule does not describe SnO$_2$ well across the entire temperature range, which can be seen in both the difference in parameters fit for the total mobility compared to those for the individual mobilities, as well as through a direct comparison of total mobilities to those computed with Matthiessen's rule (Figure S4). This behavior has also been seen in other semiconductors.[16] While the scattering rates are assumed to be independent and can be added together, this does not mean the inverse of the mobilities can likewise be added together because of self-consistency in the solution of the BTE and state-dependent scattering rates.[16] Matthiessen's rule is only a good approximation in SnO$_2$ at high temperatures, at which phonon scattering dominates.

Next, we examine the dependence of the drift mobility on the carrier concentration. While higher doping concentrations introduce more free electrons, they also contribute more scattering by ionized donor impurities. Our results (Figure 3a) show that the high mobility values in the low-doping regime (dominated by phonon scattering) decrease by approximately 50% in the high-doping limit. We fit the carrier-concentration dependence of total mobility with the empirical expression of Caughey and Thomas:[48]

$$\mu_{\text{total}}(n) = \mu_{\min} + \frac{\mu_{\max} - \mu_{\min}}{1 + (n/n_{\text{ref}})^\beta}, \tag{4}$$

where $\mu_{\max}$ is the intrinsic mobility if phonons are the only source of scattering, $\mu_{\min}$ is the mobility when ionized impurity scattering is dominant, $\beta$ characterizes how quickly the mobility changes between the two limits, and $n_{\text{ref}}$ is the doping concentration for which the mobility value is halfway between the two extremes.[49] The fitted parameters are listed in Table SIV.

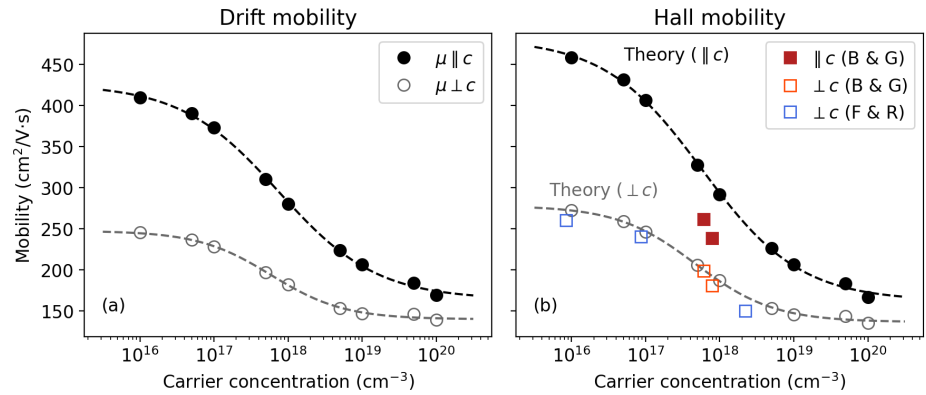

Figure 3 Room-temperature electron mobility of rutile $SnO_2$ along ∥ c and ⊥ c as a function of carrier concentration. The dashed lines are fits to the Caughey-Thomas model. (a) Drift mobility, and (b) Hall mobility compared to experimentally measured Hall values. The experimental values are from Bierwagen and Galazka (B & G) and Fonstad and Rediker (F & R).[9,12]

    Finally, to compare to experimental Hall measurements, we calculate the room-temperature electron Hall mobility of rutile $SnO_2$ as a function of carrier concentration and find good agreement with experimental values. The fitted Caughey-Thomas parameters for the Hall mobility are listed in Table SIV. Across the examined range of carrier concentrations, the Hall mobility values are similar to the drift mobility and follow the same trend (Figure 3). Experimental measurements of the Hall mobility at various carrier concentrations are also shown in Figure 3b for comparison. The agreement is excellent in the direction perpendicular to the $c$-axis, while our calculations slightly overestimate mobility compared to experiment along the parallel direction.

    We compare our results to previous theoretical investigations of carrier transport in $SnO_2$ to explore how different models and approximations affect the calculated mobility values. A study of tin-based oxide semiconductors by Hu *et al.*[13] used the RTA to calculate the phonon-limited drift and Hall mobility of $SnO_2$. Ref. 13 calculates the acoustic phonon-limited mobility using the ADP scattering model, the optical phonon-limited mobility using the Fröhlich model, and combines the two using Matthiessen's rule to obtain the total phonon-limited mobility. It finds the room-temperature drift mobility parallel to the $c$-axis to be 229 cm$^2$/V·s and perpendicular to be 166 cm$^2$/V·s. The anisotropy in these mobility values is lower than the anisotropy in our mobility values, but do agree with the anisotropy of our effective masses. This is likely because Ref. 13 assumes isotropic phonon scattering and only considers anisotropy in the directionally-dependent effective masses. The mobility values in Ref. 13 are lower than ours at nondegenerate

concentrations (below $10^{19}$ cm$^{-3}$), despite the fact that it does not consider scattering by ionized impurities. This difference could be attributed to the use of the RTA, which underestimates mobilities compared to iterative solutions of the BTE. The RTA sums scattering processes into a scattering rate that only accounts for scattering out of a given state, while iteration allows for scattering back into a state that increases the lifetime and consequently the mobility.[50] In fact, we find that the room-temperature phonon-limited mobility calculated using the RTA is 40% lower than iterative solutions of the BTE (Figure S5). Ref. 13 also calculates the Hall mobility to compare to experiment, finding a room-temperature Hall factor of 1.38 which is higher than our Hall factors that range between 0.98 and 1.12 across the temperatures and carrier concentrations (Figure S6). This discrepancy could be due to the methodology differences for computing the Hall factor. Ref. 13 uses energy-averaged scattering times that assume parabolic bands and isotropy, while our work uses the BTE augmented with a magnetic field term, which takes into account the full details of the band structure and scattering processes.

A computational study on mobility in SnO$_2$ that does take into account both phonon and ionized-impurity scattering was performed by Li *et al.*[14] Similarly to our work, the scattering mechanisms are combined by summing the individual scattering rates, which is more accurate than applying Matthiessen's rule to the individual mobilities. To calculate scattering rates, Ref. 14 uses the ADP and Fröhlich models for acoustic and polar-optical phonons, respectively, and the Brooks-Herring model for ionized impurities. The mobility is calculated with the RTA using the relaxation times from the summed scattering rates. In contrast to this work, Ref. 14 obtains an average mobility by averaging along the different directions. The total average mobility Ref. 14 finds is lower than our mobilities in both directions, most likely due to the fact that, similarly to Ref. 13, this study does not use an iterative solution of the BTE.

Many experimental measurements of mobility have been reported for SnO$_2$,[6–8,10,11] but we compare our results to that of Fonstad and Rediker[9] and Bierwagen and Galazka[12] because they specify the direction of mobility measured. Since our work finds that the mobility values are strongly directionally-dependent, a proper comparison to experiment must take this anisotropy into account. The mobility is highest along the *c* direction, which implies that aligning the *c*-axis along the direction of transport would result in better performance for SnO$_2$ electronic devices. The anisotropy we find at room temperature is stronger than that found by Bierwagen and Galazka[12] through van der Pauw measurements, which can be seen by our relatively larger mobility along the *c*-axis. The overestimation of mobility can be attributed to extra sources of scattering in experiment that our theory does not take into account, such as point defects and dislocations. This implies that to optimize device performance, it would be best to focus on minimizing other sources of scattering as the intrinsic mobility of the material itself is high. Thin-film devices in particular would benefit most from a reduction in interface roughness, which accounts for the largest additional source of scattering. Overestimation of the mobility can also be due to overscreening of the electron-phonon matrix elements by DFT.[45]

In summary, we calculate the phonon and ionized-impurity-limited electron drift and Hall mobility of rutile SnO$_2$ and examine its dependence on temperature and carrier concentration. We find that the mobility along the *c*-axis is approximately 1.6 times higher than in the perpendicular direction. The low-frequency polar-optical phonon modes contribute the most to scattering near room temperature, while high-frequency polar-optical modes dominate at higher temperatures. We also examine the carrier concentration dependence of the mobility, including the effects of both phonon and ionized-impurity scattering, and find that the trend is described well by the empirical Caughey-Thomas model observed in other semiconductors. We also find good agreement between our calculated Hall mobility values and experimental measurement, particularly along the direction perpendicular to the *c*-axis. Our work provides rigorous, first-principles insights into the dominant scattering mechanisms and the upper bounds to the electron mobility in rutile $SnO_2$ and can guide the further development of $SnO_2$-based electronic devices.

**Supplementary Material**

See the supplementary material for more details on the calculation parameters and convergence, the full band structure calculated with the LDA functional and with $G_0W_0$, the phonon dispersion and frequencies, the quadrupole tensor, the highest mode-resolved self-energies, the fitted mobility parameters, a comparison of mobilities calculated with Matthiessen's rule, a comparison of the RTA and BTE mobilities, and the Hall factors.

**Acknowledgements**

The work is supported as part of the Computational Materials Sciences Program funded by the U.S. Department of Energy, Office of Science, Basic Energy Sciences under Award No. DE-SC0020129. Computational resources were provided by the National Energy Research Scientific Computing (NERSC) Center, a DOE Office of Science User Facility, under Contract No. DEAC02–05CH11231. A.W. is supported by the Department of Defense (DoD) through the National Defense Science & Engineering Graduate (NDSEG) Fellowship Program. K.B acknowledges the support of the U.S. Department of Energy, Office of Science, Office of Advanced Scientific Computing Research, Department of Energy Computational Science Graduate Fellowship under Award Number DE-SC0020347. N.P. gratefully acknowledges the support of the Natural Sciences and Engineering Research Council of Canada (NSERC) Postgraduate Doctoral Scholarship. S. P. acknowledges support from the Fonds de la Recherche Scientifique de Belgique (FRS-FNRS) and by the Walloon Region in the strategic axe FRFS-WEL-T.

**Author Declarations**

**Conflict of Interest**

The authors have no conflicts to disclose.

**Data Availability**

The data that support the findings of this study are available from the corresponding author upon reasonable request.

# Supplementary Material for:
# Electron mobility of SnO$_2$ from first principles

Amanda Wang[1,a], Kyle Bushick[1], Nick Pant[1,2], Woncheol Lee[3], Xiao Zhang[1], Joshua Leveillee[4,5], Feliciano Giustino[4,5], Samuel Poncé[6,7], Emmanouil Kioupakis[1]

[1] Department of Materials Science and Engineering, University of Michigan, Ann Arbor, Michigan 48109, USA

[2] Applied Physics Program, University of Michigan, Ann Arbor, Michigan 48109, USA

[3] Department of Electrical and Computer Engineering, University of Michigan, Ann Arbor, Michigan 48109, USA

[4] Oden Institute for Computational Engineering and Sciences, The University of Texas at Austin, Austin, Texas 78712, USA

[5] Department of Physics, The University of Texas at Austin, Austin, Texas 78712, USA

[6] European Theoretical Spectroscopy Facility, Institute of Condensed Matter and Nanosciences, Université catholique de Louvain, Chemin des Étoiles 8, B-1348 Louvain-la-Neuve, Belgium

[7] WEL Research Institute, Avenue Pasteur, 6, 1300 Wavre, Belgium

[a] Corresponding author: amandw@umich.edu

## S1. Calculation parameters

The DFT and DFPT calculations to obtain the electronic wave functions and phonon properties used valence pseudopotentials that contain the Sn 5$s$, 5$p$, and 4$d$ electrons and the O 2$s$ and 2$p$ electrons. We relax the structure to prevent any imaginary phonon frequencies and find lattice parameters to be $a$ = 4.788 Å and c = 3.248 Å, which underestimate experimental lattice constants by 1.0% and 1.8%, respectively.[1] The total energies of the DFT calculations were converged to within 1 meV/atom with an 8×8×12 **k**-grid and 140 Ry energy cutoff.

The DFT starting point for the G$_0$W$_0$ calculations used semicore pseudopotentials, which include the Sn 4$s$ and 4$p$ electrons, because it has been shown that including semicore states can improve the agreement of the bandgap with experiment.[2,3] Because the localized semicore states would need a high energy cutoff to converge a structural relaxation, we use experimental lattice parameters in the DFT starting point calculations. The generalized plasmon-pole model was used to include the frequency dependence of the dielectric screening,[4] and the convergence of the self-energies was accelerated with respect to the sum over empty bands using the static-remainder approach.[5] The G$_0$W$_0$ calculations were converged within 20 meV error with a 45 Ry energy cutoff and a sum over 2000 and 2506 total bands for the dielectric matrix and self-energy calculations, respectively.

The EPW calculations of the mobilities were converged to within 5% with respect to the coarse grid, fine grid, and the energy window. The quasiparticle energies from the G$_0$W$_0$ calculation and the electron-phonon couplings were interpolated from a coarse 8×8×12 BZ sampling grid to a fine grid of 80×80×120 for both the **k**- and **q**-points. The number of electron states considered was truncated within an energy window around the conduction band minimum that ranged from 200 meV for the lowest carrier concentration to 500 meV for the highest carrier concentration.

## S2. Linearized BTE for drift and Hall mobilities

Carrier mobility $\mu$ is the change in carrier drift velocity with respect to electric field at the limit of zero field $\left(\mu = (dv^{drift}/dE)|_{E=0}\right)$. This is calculated from first principles by integrating over band velocities weighted by the response of the carrier occupation to electric field[6]:

$$\mu_{\alpha\beta} = \frac{-1}{V_{uc} n_c} \sum_n \int \frac{d\mathbf{k}}{\Omega_{BZ}} v_{n\mathbf{k}\alpha} \partial_{E_\beta} f_{n\mathbf{k}}, \quad (S1)$$

where α and β are Cartesian directions, $\mathbf{v}_{n\mathbf{k}}$ is the velocity of the state at band index $n$ and crystal momentum $\mathbf{k}$, and $\partial_{E_\beta} f_{n\mathbf{k}} \equiv (\partial f_{n\mathbf{k}}/\partial E_\beta)|_{E=0}$ is the linear response of the carrier occupation with respect to the electric field. The drift mobility, or the response of the carrier drift velocity in the presence of only an electric field, is calculated by solving for the linear response coefficients using the linearized BTE:

$$\partial_{E_\beta} f_{n\mathbf{k}} = e v_{n\mathbf{k}} \frac{\partial f^0_{n\mathbf{k}}}{\partial \varepsilon_{n\mathbf{k}}} \tau_{n\mathbf{k}} + \tau_{n\mathbf{k}} \sum_{m\mathbf{q}} \Gamma_{m\mathbf{k}+\mathbf{q} \to n\mathbf{k}} \partial_{E_\beta} f_{m\mathbf{k}+\mathbf{q}}. \quad (S2)$$

This equation has the linear response coefficients on both the left- and right-hand sides and is solved iteratively until self-consistency is reached. The self-energy relaxation time approximation can be derived by dropping the last term in Eqn. S2, which removes the need for iteration. $f^0_{n\mathbf{k}}$ is the equilibrium Fermi-Dirac occupation and $\varepsilon_{n\mathbf{k}}$ is the energy of the $n\mathbf{k}$-th state. $\tau_{n\mathbf{k}}$ is the state-dependent total scattering lifetime and its inverse is defined as:

$$\tau^{-1}_{n\mathbf{k}} = \sum_{m\mathbf{q}} \Gamma_{n\mathbf{k} \to m\mathbf{k}+\mathbf{q}}. \quad (S3)$$

$\Gamma_{n\mathbf{k} \to m\mathbf{k}+\mathbf{q}}$ is the partial transition rate from the $n\mathbf{k}$-th state to the $m\mathbf{k}+\mathbf{q}$-th state and, in this work, is the sum of transitions due to electron-phonon scattering and electron-ionized impurity scattering:

$$\Gamma_{n\mathbf{k} \to m\mathbf{k}+\mathbf{q}} = \Gamma^{ph}_{n\mathbf{k} \to m\mathbf{k}+\mathbf{q}} + \Gamma^{ii}_{n\mathbf{k} \to m\mathbf{k}+\mathbf{q}}. \quad (S4)$$

The partial transition rate due to electron-phonon scattering is calculated by summing over all phonon modes, using the electron-phonon matrix elements.[7] The partial transition rate due to electron-ionized impurity scattering is calculated using an ensemble average of randomly distributed point charges.[8]

To calculate Hall mobility, or the response of the carrier drift velocity in the presence of both an electric and a magnetic field, the linearized BTE is augmented with a magnetic field term:

$$\left[1 - \frac{e}{\hbar} \tau_{n\mathbf{k}} (v_{n\mathbf{k}} \times \mathbf{B}) \cdot \nabla_\mathbf{k}\right] \partial_{E_\beta} f_{n\mathbf{k}} = e v_{n\mathbf{k}} \frac{\partial f^0_{n\mathbf{k}}}{\partial \varepsilon_{n\mathbf{k}}} \tau_{n\mathbf{k}} + \tau_{n\mathbf{k}} \sum_{m\mathbf{q}} \Gamma_{m\mathbf{k}+\mathbf{q} \to n\mathbf{k}} \partial_{E_\beta} f_{m\mathbf{k}+\mathbf{q}}. \quad (S5)$$

A small, finite magnetic field of $10^{-10}$ T is applied in the calculation and the gradient calculated with finite differences. The linear response coefficients from the self-consistent solution of this equation are used to calculate the Hall mobility.

## S3. Band structure

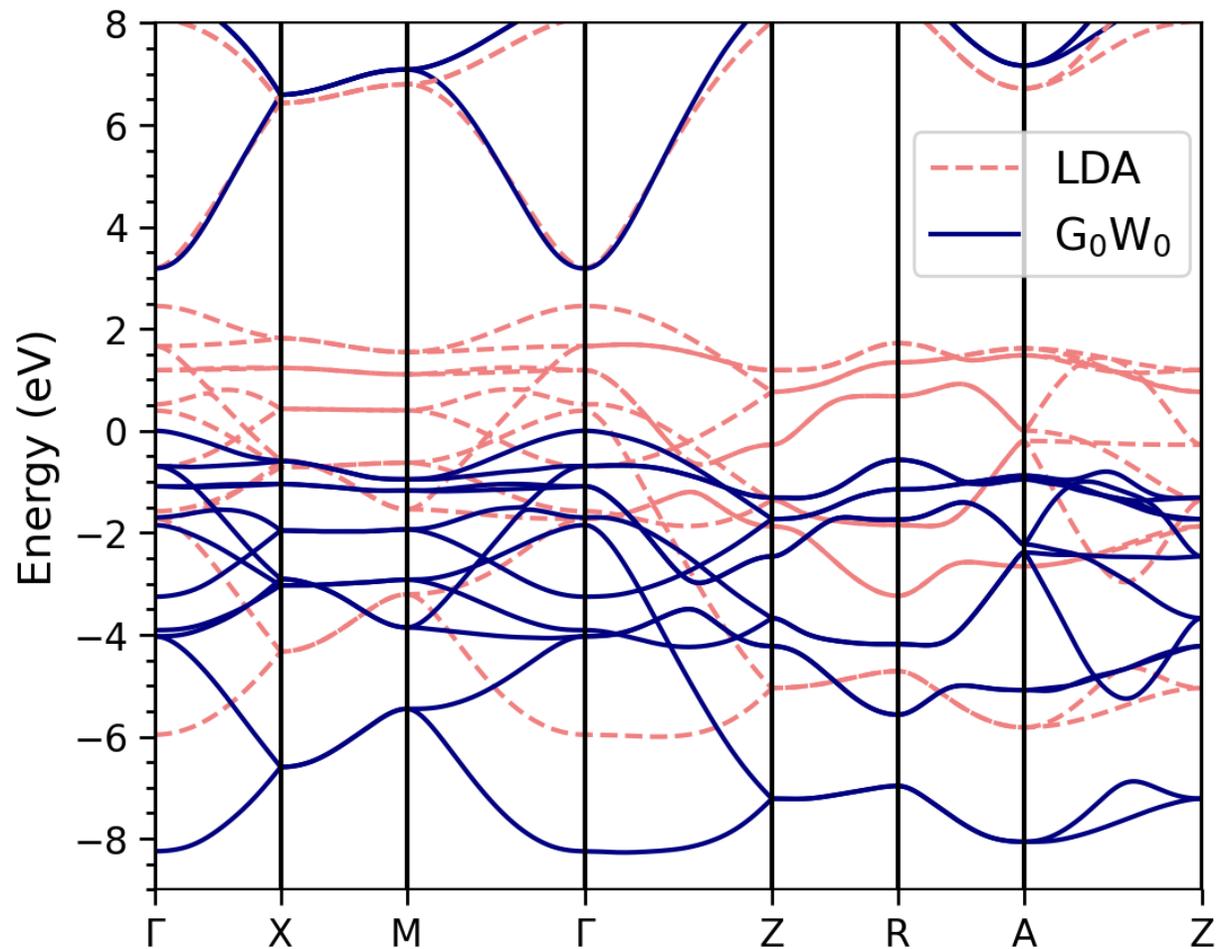

*Figure S1 $G_0W_0$ quasiparticle band structure (solid) and LDA band structure (dotted) of rutile tin dioxide along selected high-symmetry Brillouin-zone paths. The bands are aligned at the conduction band minimum, which is located at Γ. The direct bandgap is located at Γ with a magnitude of 3.182 eV ($G_0W_0$) and 0.749 eV (LDA).*

## S4. Phonon properties

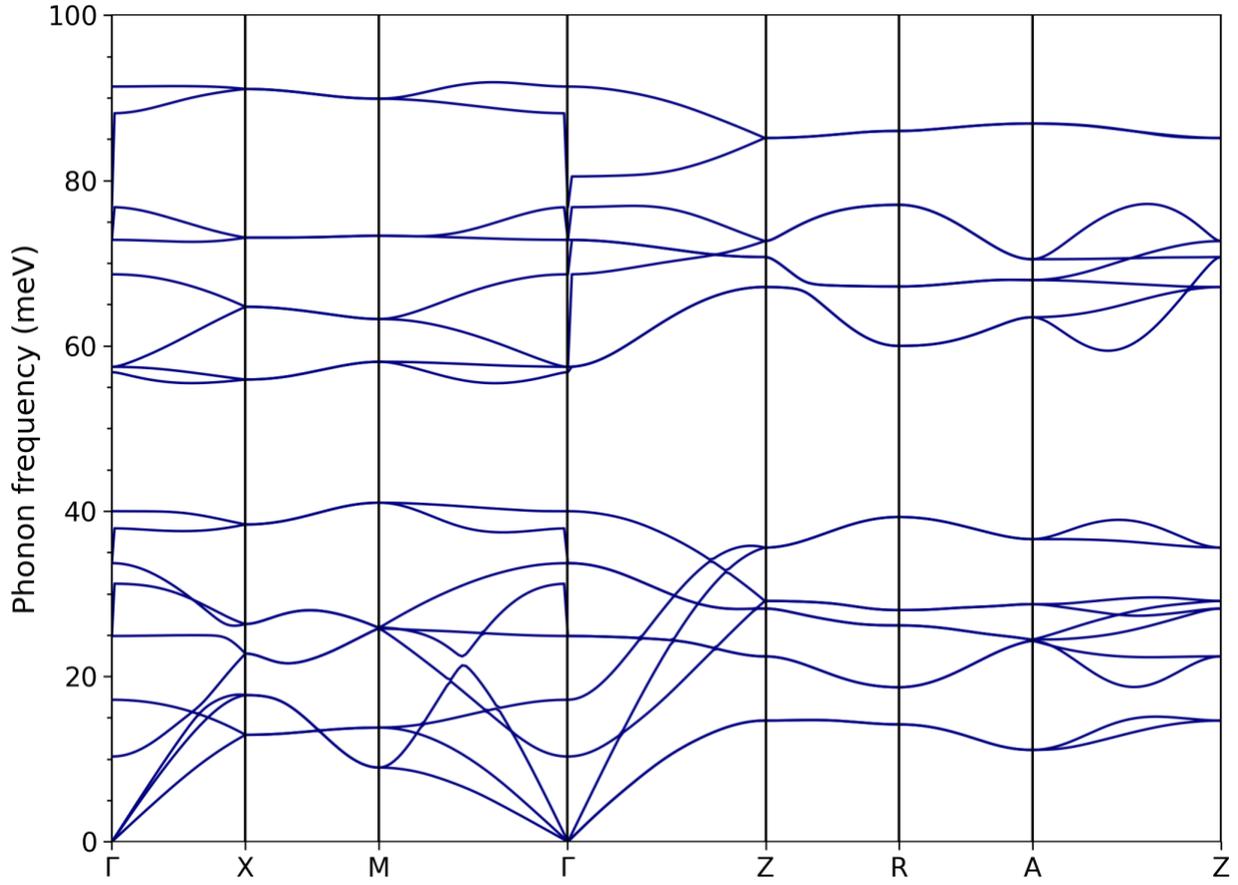

*Figure S2 Phonon dispersion of rutile tin dioxide along selected high-symmetry Brillouin-zone paths.*

Table SI Optical phonon frequencies (meV) at the Brillouin zone center of rutile tin dioxide calculated in this work using the LDA functional, compared with values from experimental studies and previous theoretical studies also using the LDA functional.

| Phonon mode | This work (theory) | Previous theory[9] | Experiment[10] |
|---|---:|---:|---:|
| $B_{1g}$ | 10.3 | 10.2 | – |
| $B_{1u}$ | 17.2 | 17.2 | – |
| $E_u$ | 24.9 | 24.8 | 30.3 |
| $E_u$ | 31.2 | 31.2 | 34.2 |
| $E_u$ | 33.7 | 33.5 | 36.3 |
| $E_u$ | 37.9 | 38.0 | 45.4 |
| $A_{2g}$ | 39.9 | 39.7 | – |
| $A_{2u}$ | 56.8 | 56.7 | 59.1 |
| $E_g$ | 57.4 | 57.3 | 59.0 |
| $B_{1u}$ | 68.7 | 68.5 | – |
| $E_u$ | 72.8 | 72.4 | 76.6 |
| $A_{1g}$ | 76.7 | 76.5 | 79.1 |
| $A_{2u}$ | 80.5 | 80.4 | 87.4 |
| $E_u$ | 88.2 | 88.2 | 95.5 |
| $B_{2g}$ | 91.4 | 91.0 | 97.0 |

**S5. Quadrupole tensor**

The interpolation of the electron-phonon matrix elements includes both dipole and quadrupole corrections. The quadrupole corrections were included by supplying the quadrupole tensor (listed below) to the EPW calculations. The quadrupole tensor was computed using perturbation theory[11] as implemented in the ABINIT software.[12,13] The tin atoms have zero quadrupole tensor because their octahedral sites have inversion symmetry, while the oxygen atoms do not.[11] We also ran an EPW calculation without the quadrupole tensor to quantify the effect of the quadrupole correction, and we found a less than 1% difference in the mobility values. At 300 K and a carrier concentration of $10^{17}$ cm$^{-3}$, the phonon-limited mobilities are 413.29 cm²/V·s and 247.24 cm²/V·s parallel and perpendicular to the $c$-axis, respectively, with quadrupole corrections, and 413.65 cm²/V·s and 247.91 cm²/V·s without quadrupole corrections. Quadrupole corrections have been shown to mainly affect acoustic modes in polar semiconductors.[14] They likely have little effect in this work because it is the polar-optical phonons that dominate electron-phonon scattering.

Table SII The quadrupole tensor (in $e$ Bohr) of rutile tin dioxide calculated using ABINIT. Used for the interpolation of the long-range component of the electron-phonon matrix elements by EPW. ($Q_{\kappa\alpha}^{\beta\gamma} = Q_{\kappa\alpha}^{\gamma\beta}$)

| κ | α | $Q^{11}$ | $Q^{22}$ | $Q^{33}$ | $Q^{23}$ | $Q^{13}$ | $Q^{12}$ |
|---|---|---|---|---|---|---|---|
| Sn$_1$ | 1 | 0 | 0 | 0 | 0 | 0 | 0 |
| | 2 | 0 | 0 | 0 | 0 | 0 | 0 |
| | 3 | 0 | 0 | 0 | 0 | 0 | 0 |
| Sn$_2$ | 1 | 0 | 0 | 0 | 0 | 0 | 0 |
| | 2 | 0 | 0 | 0 | 0 | 0 | 0 |
| | 3 | 0 | 0 | 0 | 0 | 0 | 0 |
| O$_1$ | 1 | 2.44970 | 1.32250 | -0.37641 | 0 | 0 | 0.02175 |
| | 2 | -1.32250 | -2.44970 | 0.37641 | 0 | 0 | -0.02175 |
| | 3 | 0 | 0 | 0 | 0.21113 | -0.21113 | 0 |
| O$_2$ | 1 | -2.44970 | -1.32250 | 0.37641 | 0 | 0 | 0.02175 |
| | 2 | -1.32250 | -2.44970 | 0.37641 | 0 | 0 | 0.02175 |

| | 3 | 0 | 0 | 0 | 0.21113 | 0.21113 | 0 |
| --- | --- | --- | --- | --- | --- | --- | --- |
| $O_3$ | 1 | 2.44970 | 1.32250 | -0.37641 | 0 | 0 | -0.02175 |
| | 2 | 1.32250 | 2.44970 | -0.37641 | 0 | 0 | -0.02175 |
| | 3 | 0 | 0 | 0 | -0.21113 | -0.21113 | 0 |
| $O_4$ | 1 | -2.44970 | -1.32250 | 0.37641 | 0 | 0 | -0.02175 |
| | 2 | 1.32250 | 2.44970 | -0.37641 | 0 | 0 | 0.02175 |
| | 3 | 0 | 0 | 0 | -0.21113 | 0.21113 | 0 |

## S6. Electron self-energies

Labeling the phonon modes according to increasing frequency at the BZ center, the $E_u$ modes 7, 9, and 17 have the largest imaginary self-energies, with mode 17 showing a sharp increase at approximately 80 meV above the conduction band where there is enough energy for phonon emission. Modes 7, 9, and 17 have frequencies of 31.2, 37.9, and 88.2 meV, respectively, which correspond well to both the range of characteristic temperatures that we fit for the temperature dependence of the phonon-limited mobility (Table SIII) and the peaks in the spectral decomposition of the electron-phonon scattering rates (Figure 2b).

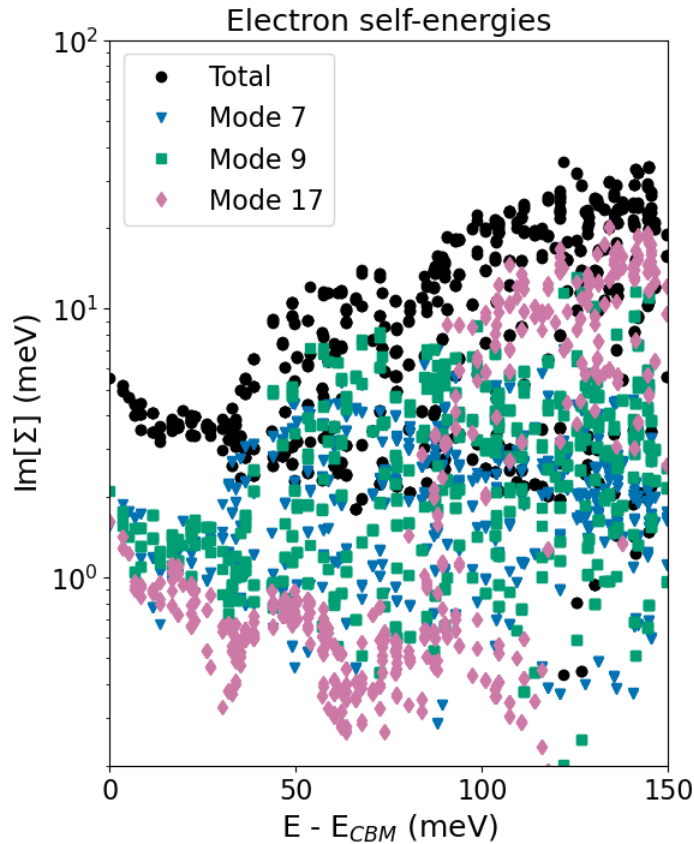

*Figure S3 Imaginary part of the electron self-energies due to the electron-phonon interaction. The total self-energy is shown along with mode-resolved energies of the top three modes.*

## S7. Mobility parameters

We fit the temperature and carrier concentration dependence of the electron mobility to the analytical models described in the main text. Listed here are the parameters of the models.

Table SIII Fitted parameters for the temperature dependence of the phonon-limited, ionized impurity-limited, and total drift mobilities.

|  | $\mu_{\text{low}}$ (cm$^2$/V·s) | $\mu_{\text{high}}$ (cm$^2$/V·s) | $T_{\text{low}}$ (K) | $T_{\text{high}}$ (K) | $\alpha$ |
|---|---|---|---|---|---|
| $\mu_{\text{ph}}(T) \parallel c$ | 228.12 | 34.96 | 361.10 | 972.23 | - |
| $\mu_{\text{ph}}(T) \perp c$ | 114.50 | 33.79 | 393.89 | 854.67 | - |
| $\mu_{\text{ii}}(T) \parallel c$ | - | - | - | - | 1.01 |
| $\mu_{\text{ii}}(T) \perp c$ | - | - | - | - | 1.07 |
| $\mu_{\text{total}}(T) \parallel c$ | 713.44 | 36.73 | 123.23 | 827.82 | 0.0 |
| $\mu_{\text{total}}(T) \perp c$ | 406.41 | 30.15 | 146.18 | 736.88 | 0.0 |

Table SIV Fitted parameters for the carrier concentration dependence of the total drift and Hall mobility.

|  | $\mu_{\text{min}}$ (cm$^2$/V·s) | $\mu_{\text{max}}$ (cm$^2$/V·s) | $\beta$ | $n_{\text{ref}}$ (cm$^{-3}$) |
|---|---|---|---|---|
| $\mu_{\text{total}}(n) \parallel c$ | 163.80 | 425.89 | 0.656 | 7.49×10$^{17}$ |
| $\mu_{\text{total}}(n) \perp c$ | 139.72 | 247.26 | 0.879 | 5.90×10$^{17}$ |
| $\mu_{\text{total}}^{\text{Hall}}(n) \parallel c$ | 161.99 | 481.16 | 0.657 | 5.82×10$^{17}$ |
| $\mu_{\text{total}}^{\text{Hall}}(n) \perp c$ | 136.33 | 277.77 | 0.820 | 4.72×10$^{17}$ |

## S8. Matthiessen's rule

Matthiessen's rule is a widely used approximation that mobilities limited by different scattering mechanisms add in inverse[15]:

$$\frac{1}{\mu_{total}} = \frac{1}{\mu_{ph}} + \frac{1}{\mu_{ii}}.$$

In contrast, we calculate total mobilities by summing the transition rates due to the various scattering mechanisms and then solving the BTE with the total transition rates. A comparison of the total mobilities calculated each way is shown below in Figure S4. While Matthiessen's rule does describe the total mobility well in the high temperature regime at which phonon scattering dominates, it overestimates mobility at lower temperatures where electron-phonon and electron-ionized impurity scattering are comparable.

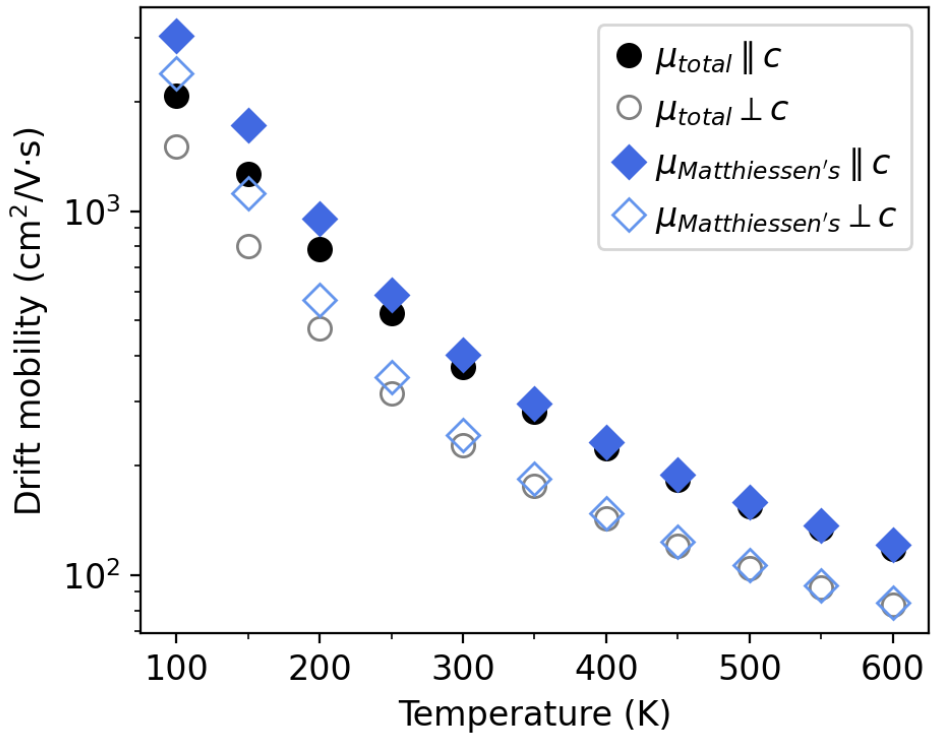

*Figure S4 Total mobilities computed by summing the scattering rates and then solving the BTE compared to mobilities computed using Matthiessen's rule to combine the phonon-limited and ionized-impurity-limited mobilities.*

## S9. SERTA vs IBTE mobility

For the highest level of accuracy, the Boltzmann transport equation needs to be solved self-consistently to find the linear response of the carrier distribution to the electric field, which can then be used to calculate mobility. The self-consistent procedure is often referred to as the iterative Boltzmann transport equation (IBTE). A faster but less accurate approach that doesn't require iteration is the self-energy relaxation time approximation (SERTA), which is a specific form of the relaxation time approximation.[7] The relaxation time approximation often leads to underestimated mobilities,[16] which we observe in this system.

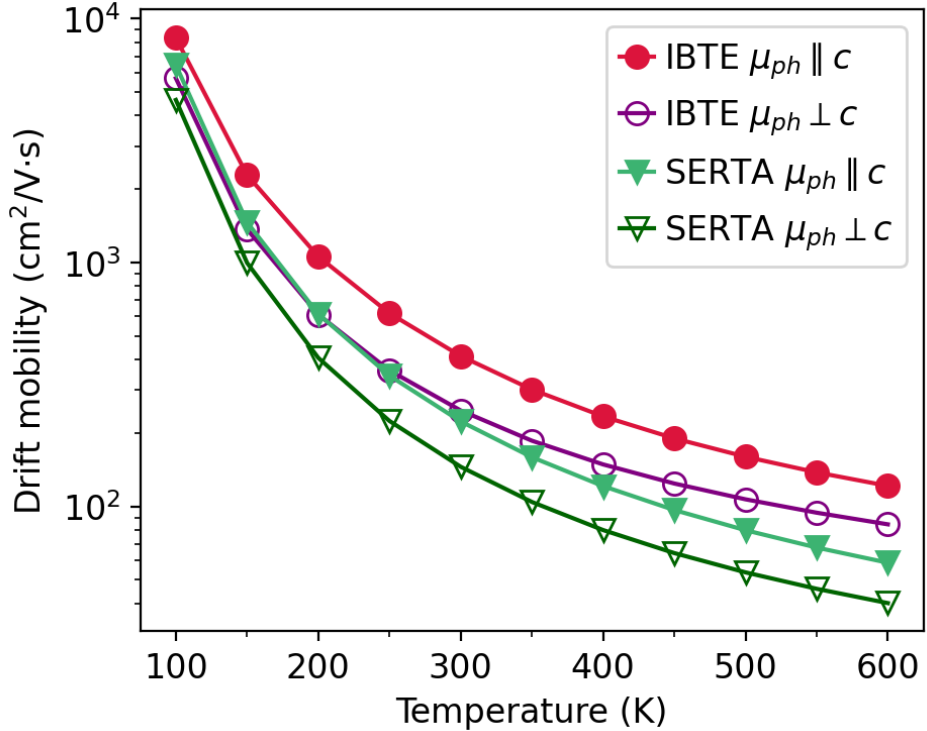

*Figure S5 The phonon-limited mobilities calculated from iterative solutions of the Boltzmann transport equation (IBTE) and from the self-energy relaxation time approximation (SERTA) as a function of temperature.*

## S10. Hall factors

To calculate Hall mobility, we solve a form of the BTE augmented with an electric field term.[17] We find similar drift mobilities and Hall mobilities, signified by Hall factors close to 1 across the range of different temperatures and carrier concentrations.

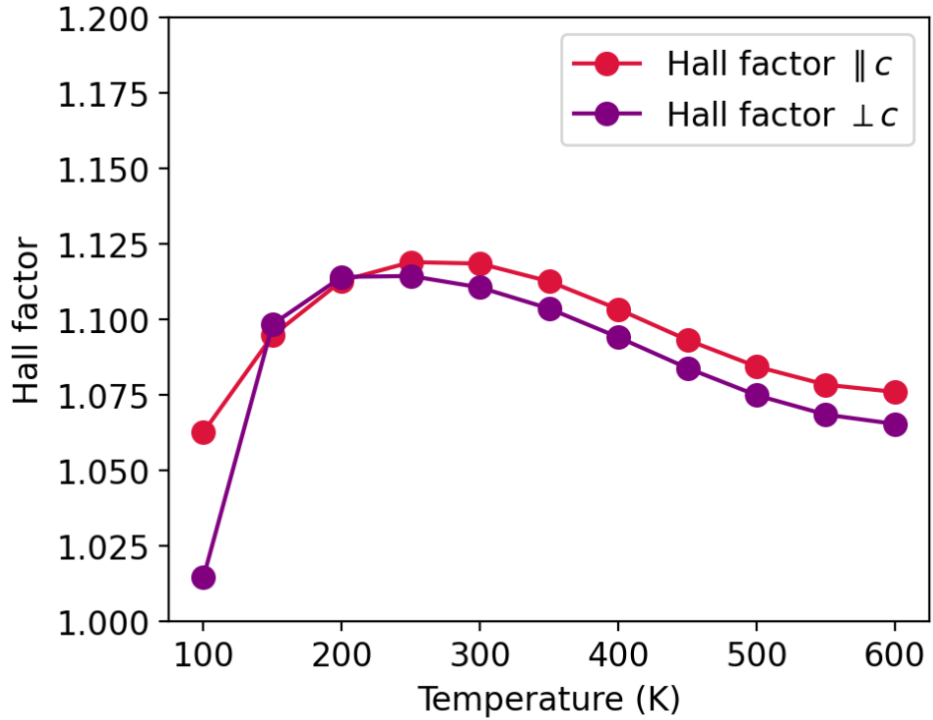

*Figure S6 The Hall factors for the phonon-limited mobilities as a function of temperature.*